\documentclass[pr,twocolumn,preprintnumbers,nofootinbib,showpacs,amsmath,amssymb,superscriptaddress]{revtex4-1}
\usepackage{ifpdf}
\usepackage{amsfonts}
\usepackage{mathrsfs}
\usepackage{amssymb}
\usepackage{amsmath}
\usepackage{color}
\usepackage{amsfonts}
\usepackage{graphicx}
\usepackage{epstopdf}
\usepackage[
      colorlinks=true,
      linkcolor=blue,
      urlcolor=blue,
      filecolor=black,
      citecolor=blue,
            ]{hyperref}

\begin{document}
\title{A Holographic Model for Paramagnetism/antiferromagnetism Phase Transition}
\author{Rong-Gen Cai}
\email{cairg@itp.ac.cn}
\author{Run-Qiu Yang}
\email{aqiu@itp.ac.cn}
\affiliation{State Key Laboratory of Theoretical Physics,Institute of Theoretical Physics,\\
 Chinese Academy of Sciences,Beijing 100190, China}

\begin{abstract}
In this paper we build a holographic model of paramagnetism/antiferromagnetism phase transition, which is realized by introducing two real antisymmetric tensor fields coupling to the background gauge field strength and interacting with each other in a dyonic black brane background. In the case without external magnetic field and in low temperatures, the magnetic moments condense spontaneously in antiparallel manner with the same magnitude and the time reversal symmetry is also broken spontaneously (if boundary spatial dimension is more than 2, spatial rotational symmetry is broken spontaneously as well), which leads to an antiferromagnetic phase. In the case with weak external magnetic field, the magnetic susceptibility density has a peak at the critical temperature and satisfies the Curie-Weiss law in the paramagnetic phase of antiferromagnetism. In the strong external magnetic field case, there is a critical magnetic field $B_c$ in antiferromagnetic phase: when magnetic field reaches $B_c$, the system will return into the paramagnetic phase by a second order phase transition.
\end{abstract}
\maketitle

\noindent

\section{Introduction}
\label{intro}
The anti-de Sitter/conformal field theory (AdS/CFT) correspondence relates a weakly coupled gravity theory in anti-de Sitter (AdS) space  to a strongly coupled conformal field theory (CFT) living on the AdS boundary~\cite{Maldacena:1997re,Gubser:1998bc,Witten:1998qj,Witten:1998qj2}.
Due to the existence of scaling symmetry near critical point, the AdS/CFT correspondence provides
a powerful tool to study critical phenomena in some strongly coupled condensed matter systems. This duality also gives us a way to understand gravity and condensed matter physics from other side. Over the past years there have been a lot of studies of the applications of the AdS/CFT duality in condensed matter physics (for reviews, see Refs.~\cite{Hartnoll:2009sz,Herzog:2009xv,McGreevy:2009xe,Horowitz:2010gk,Cai:2015cya}). Recently, some efforts have been made to generalize the correspondence to systems with less symmetries (see Refs.~\cite{Nakamura:2009tf,Donos:2011bh,Horowitz:2012ky,Horowitz:2013jaa,Ling:2013aya,Rozali:2013ama,Cai:2013sua}, for example) and to the far from thermal equilibrium problems (see Refs.~\cite{Murata:2010dx,Bhaseen:2012gg,Adams:2012pj,Garcia-Garcia:2013rha,Chesler:2013lia}, for example).

In a previous paper~\cite{Cai:2014oca}, the present authors proposed a new example of the application of the AdS/CFT correspondence, by
realizing the holographic description of the paramagnetism/ferromagnetism phase transition in a dyonic Reissner-Nordstr\"om-AdS black brane. In that model,  the magnetic moment is realized  by condensation of a real antisymmetric tensor field which couples to the background gauge field strength in the bulk. In the case without external magnetic filed, the time reversal symmetry is broken spontaneously and spontaneous magnetization happens in low temperatures. The critical exponents are in agreement with the ones from mean field theory. In the case of nonzero magnetic field, the model realizes the hysteresis loop of single magnetic domain and the magnetic susceptibility satisfies the Curie-Weiss law.

Except for the paramagnetism and ferromagnetism, antiferromagnetism is another important magnetic property of material. So a natural question is whether we can extend the model in Ref.~\cite{Cai:2014oca} to realize antiferromagnetism. To do that, let us first
briefly review some characteristic properties and some relevant theory of antiferromagnetism~\cite{A.B.L,C.Kittel}.

\begin{figure}[h!]
\begin{center}
\includegraphics[width=0.23\textwidth]{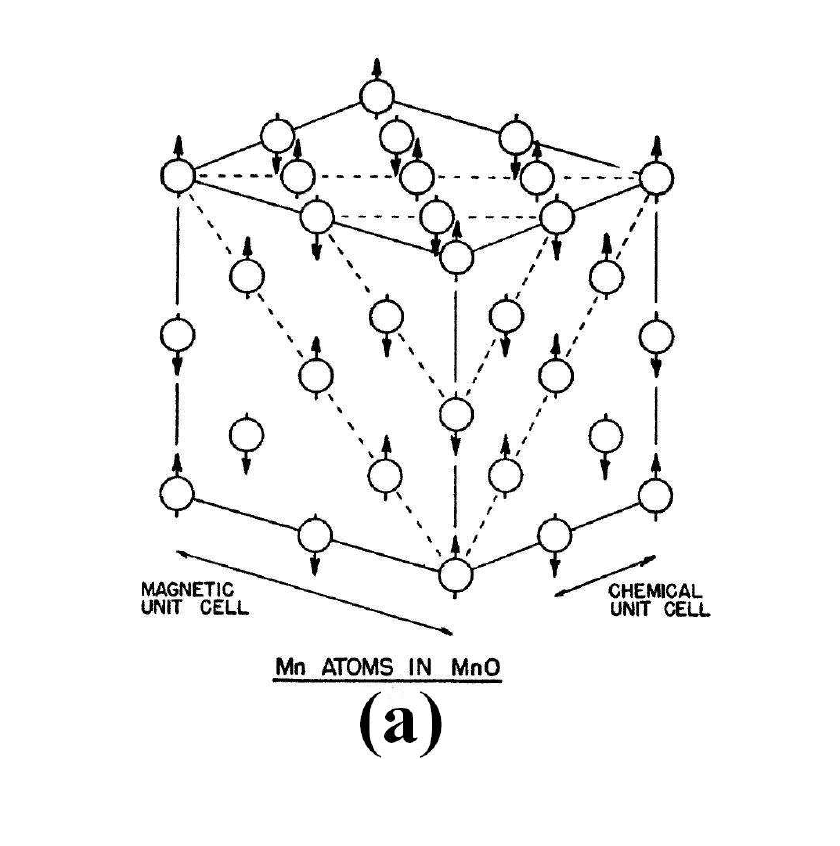}
\includegraphics[width=0.25\textwidth]{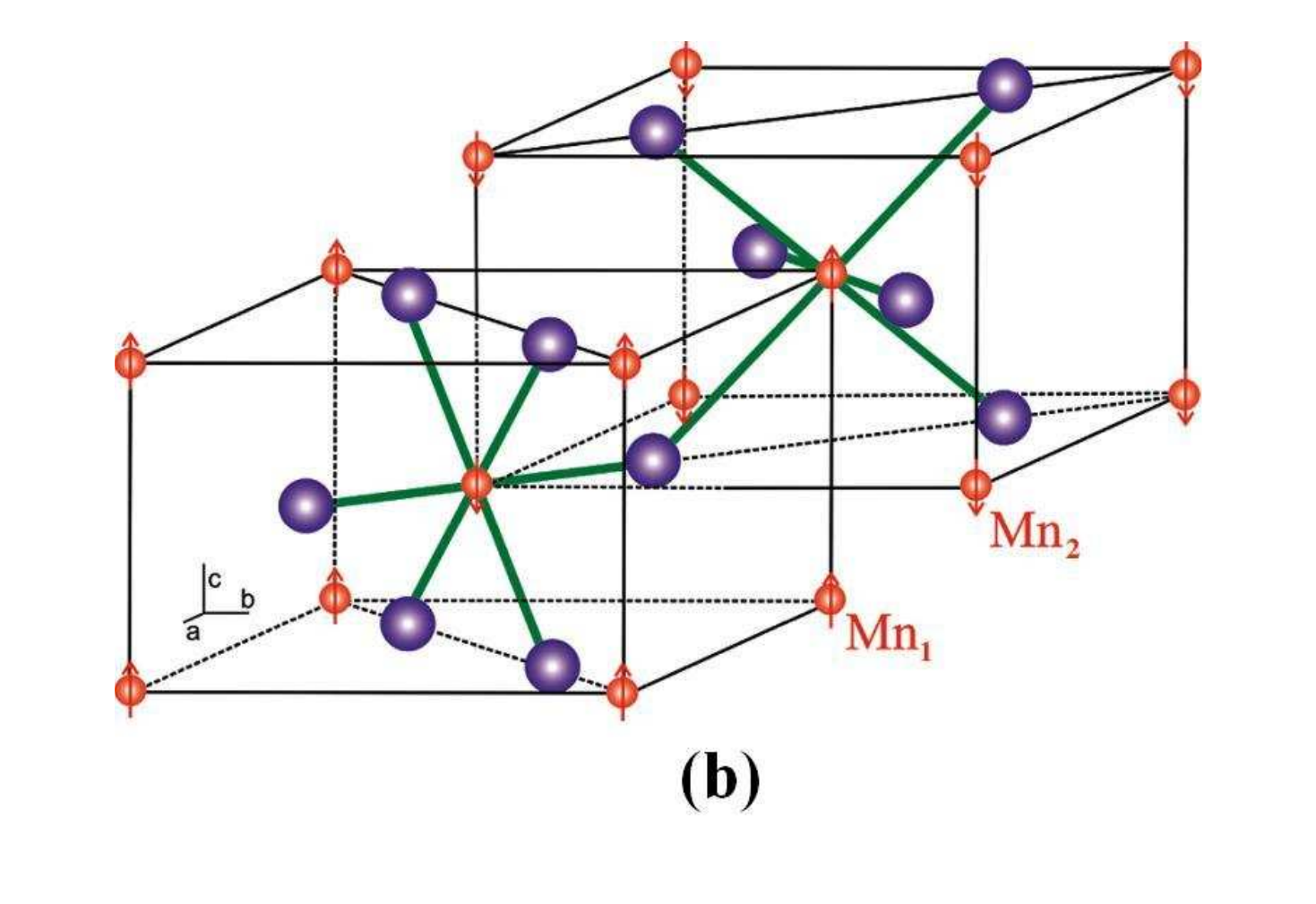}
\caption{ Schematic representation of spin alignments for antiferromagnetics at 0 K. (a) Display of magnetic structure in MnO. The arrows stands for the directions of spin and circles represent the Mn ions~\cite{g.G.S1}. The oxygen ions (are not shown here) do not contribute to the antiferromagnetic behavior. MnO has a NaCl structure. (b) Three-dimensional representation of the spin alignment of manganese ions in MnF$_2$~\cite{z.Yaman1}. The smaller spheres stand for Mn ions and larger ones stand for fluoride ions. This figure demonstrates the interpenetration of two manganese sub-lattices, Mn$_1$ and Mn$_2$, having antiparallel aligned moments.}
\end{center}
\label{TM2}
\end{figure}

Though an antiferromagnetic material does not show any macroscopic magnetic moment when external magnetic field is absent, it is still a kind of magnetic ordered material when temperature is below the N\'{e}el temperature $T_N$. The conventional picture, due to L. N\'{e}el, represents a macroscopic antiferromagnetism as consisting of two sublattices~\footnote{There are some antiferromagnetic materials which have more than two sublattices. In those materials, the spins in the sublattices are arranged with zero total spin. These kinds of materials can be regarded as the extension of the two sublattices. In this paper, as a toy model, we will considere the case with two sublattices only.}, call them A and B such that spins on one sublattice point opposite to those of the other sublattice~\footnote{There are some antiferromagnetic materials, in which the magnetic moments are a little canted with each other near the transition temperature, such as $\alpha$-Fe$_2$O$_3$ in the temperature of 950-260K. This phenomenon originates from some scattering and perturbation effects of spin-orbit coupling~\cite{T.Moriya}. However, when temperature approaches to zero, the magnetic moments become antiparallel. In this paper, we will not consider these cases. }, i.e., $\overrightarrow{M}_A=-\overrightarrow{M}_B$. The order parameter is the staggered magnetization, defined as the difference between the two magnetic moments associated with the two sublattices:
\begin{equation}\label{stagg1}
\overrightarrow{M}^\dagger=\overrightarrow{M}_A-\overrightarrow{M}_B.
\end{equation}
Such a complicated ordering is not straightforward to identify by thermodynamic measurements, and in early days even the existence of antiferromagnetism was hotly debated. However, the technology of neutron diffraction gives us a powerful tool to directly measure the microscopic magnetic moment distribution of the lattices. The experiments confirmed this picture in the antiferromagnetic insulator.  Figure~\ref{TM2} depicts the spin alignments for two manganese compounds. (Only the spins of the manganese ions contribute to the antiferromagnetic behavior.) Figure~\ref{TM2} (a) shows that the ions in a given \{111\} plane possess parallel spin alignment, whereas ions in the adjacent plane have antiparallel spins with respect to the first plane. In Figure~\ref{TM2} (b), the spins of manganese ions in vertexes are all parallel alignment, whereas ions in the body-centered are arranged in antiparallel manner with vertexes. Thus, both in these two materials, the magnetic moments of the solid cancel each other and the material as a whole has no net magnetic moment.

One can see that, below the N\'{e}el temperature, the magnetic moment in antiferromagnetic material is still ordered rather than the case in paramagnetism where the magnetic moments are arranged randomly because of thermal fluctuations. However, when the temperature increases  beyond $T_N$, the ordered magnetic moment structure is broken and the magnetic moment begins to distribute randomly, which corresponds to the paramagnetic phase of the antiferromagnetic material.

\begin{figure}[h!]
\includegraphics[width=0.35\textwidth]{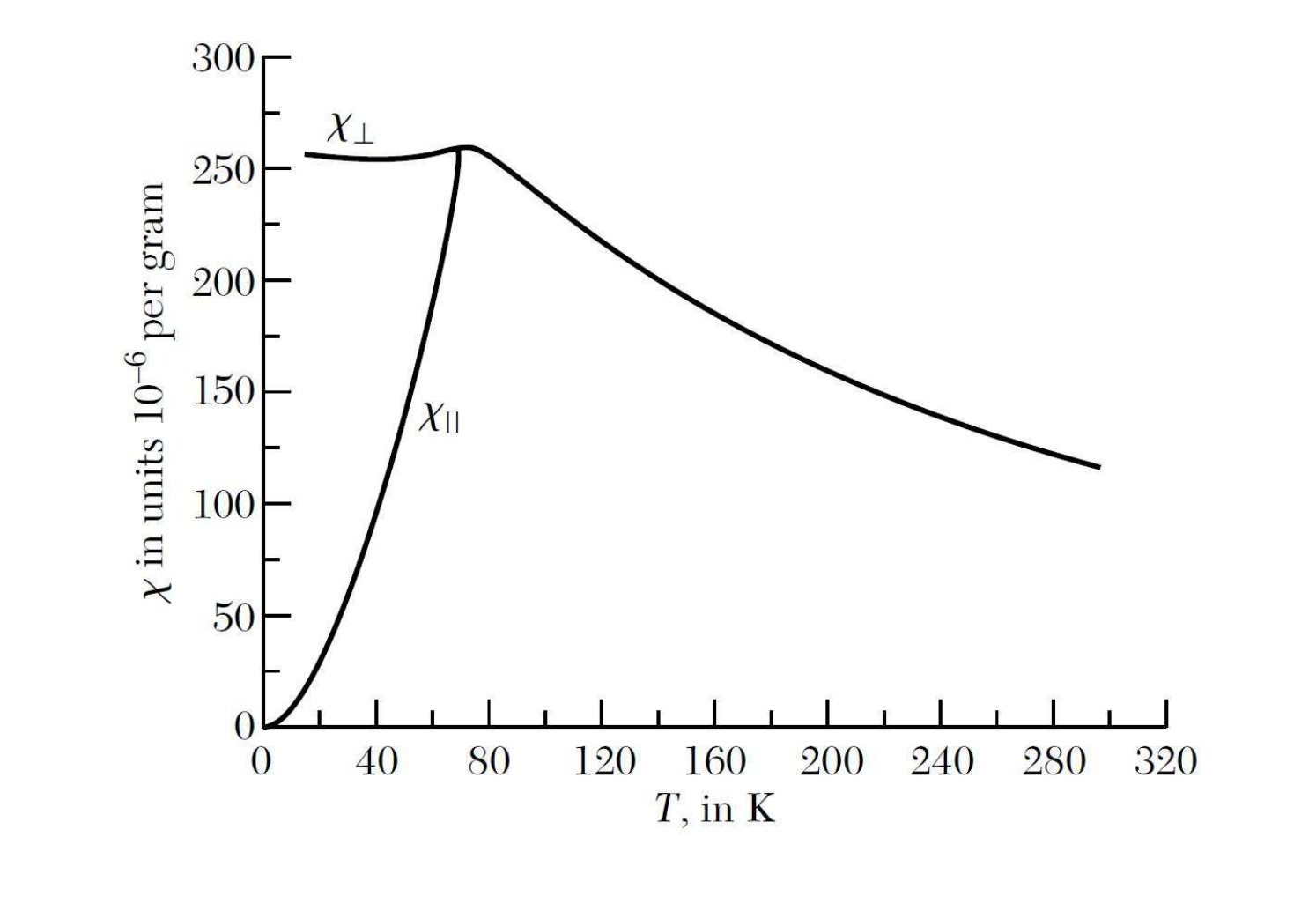}
\includegraphics[width=0.2\textwidth]{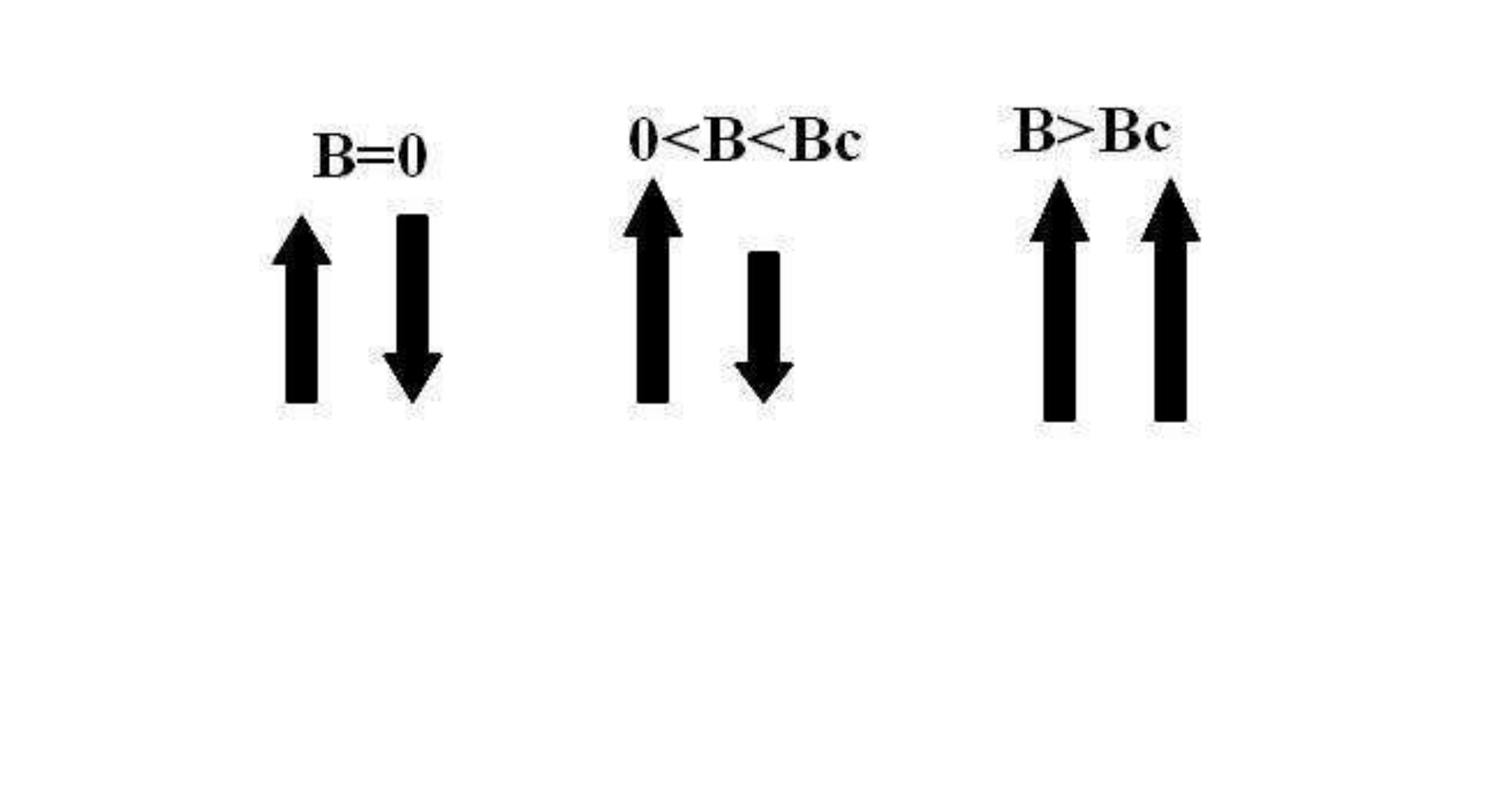}
\caption{(Top panel) The magnetic susceptibility of antiferromagnetic material MnF$_2$~\cite{C.Kittel}. $\chi_{\perp}$ stands for the case with external magnetic field  perpendicular to the direction of spontaneous magnetization and $\chi_{\parallel}$ for the case with external magnetic field parallel to the direction of spontaneous magnetization. (Bottom panel) The influence of external magnetic field on the magnetic moments of two sublattices when $T<T_N$. }
\label{TM3}
\end{figure}

There is another significant difference between the antiferromagnetism and paramagnetism which can be seen from  magnetic susceptibility. The behavior of magnetic susceptibility of antiferromagnetism  is shown schematically in figure~\ref{TM3}. In the high temperature region $T>T_N$, the antiferromagnetic material is in a paramagnetic phase whose magnetic moments are randomly distributed. So the susceptibility performs as a paramagnetism and obeys the Curie-Weiss law~\cite{R.M.W}:
\begin{equation}\label{chi1}
\chi=\frac C{T+\theta},\ \ \ T>T_N,\ \ \theta>0,
\end{equation}
where $C$ and $\theta$ are two constants. Note that in the paramagnetic material and the paramagnetic phase of ferromagnetic material, the magnetic susceptibility also obeys the Curie-Weiss law, but the constant $\theta$ in \eqref{chi1} is zero and negative, respectively. For the antiferromagnetic material, in the region of low temperature $T<T_N$, the magnetic moments condense and arrange into two opposite directions. In an external magnetic field, a kind of ferrimagnetic behavior may be displayed in the antiferromagnetic phase, with the absolute value of one of the sublattice magnetizations differing from that of the other sublattice, resulting in a nonzero net magnetization. For a weak  external field, there are two cases according to the directions of external magnetic field and spontaneous magnetization. When the external field is perpendicular to spontaneously induced magnetic moment, the magnetic susceptibility is almost independent of temperature, while in the parallel case, the susceptibility decreases when temperature is lowered, because the force making magnetic moments antiparallel becomes stronger and stronger, which leads to that the material is harder and harder to be magnetized by external field. However, if the external field is enough strong, it will destroy the antiferromagnetic interaction of the two magnetic moment. Then the system will perform a paramagnetism, i.e., both two sublattices have same magnetic moments paralleling the external magnetic field as if they have no interaction with each other (see the right plot of figure~\ref{TM3}).

In general, magnetism of materials originates from the exchange interaction of electrons. There are two methods to deal with the magnetic ordering in materials. One is based on the local magnetic moment and the basic model is Heisenberg Hamiltonian. The other is based on the energy band theory with the effects of spin and the basic model is Hubbard Hamiltonian. The former is suitable to describe magnetic properties in insulator and some semiconductor, while the latter is suitable for conductor. In both models, the exchange interaction plays a crucial role.
Various microscopic exchange interactions between the magnetic moments or spins may lead to antiferromagnetic structures. Generally speaking, a negative exchange interaction is the main reason for antiferromagnetism. In the simplest case, one may consider an Ising model on an bipartite lattice, e.g., the simple cubic lattice, with couplings between spins at nearest neighbor sites. Depending on the sign of the interaction, ferromagnetic or antiferromagnetic order will appear. Geometrical frustration or competing between ferro- and antiferro-magnetic interactions may lead to different and, perhaps, more complicated magnetic structures~\cite{R.M.W}. These theories  give us a very deep understanding in the weak correlation system. However, in recent years, many antiferromagnetic or ferromagnetic materials show that strong correlation plays important role and exhibits some novel properties~\cite{DLQ1,AAFM,A.P.Ramirez:1995ge}. So for these materials, the AdS/CFT correspondence might provide a useful method to understand these properties.

The typical paramagnetism/antiferromagnetism phase transition is a second ordered phase transition. So this phase transition should be associated  with some symmetry breaking spontaneously. Normally the appearance of antiferromagnetism (or ferromagnetism) is associated with the breaking of a continuous symmetry-spatial rotations and a $\mathbb{Z}_2$ symmetry-time reversal simultaneously in three dimensions\footnote{The magnetic moment is  proportional to the spins and orbital angular moment (in some case, the latter can be neglected), so it will obtain a minus sign under the time reversal transformation.}. Though the former is more obvious, the latter is a characteristic feature. Since the magnetic field and spin (or angular moment) are vectors (in fact they are axial vectors) only in three spatial dimensions. As a theoretical consideration, we can also build an antiferromagnetism (or ferromagnetism) models in two spatial dimensions\footnote{In two spatial dimension, isotropic Heisenberg model with finite-range exchange interaction can be neither ferromagnetic nor antiferromagnetic~\cite{N.D.M}. However, it does't exclude anisotropic cases or the models beyond Heisenberg model.}, where the magnetic field and spin (or angular moment) are pseudo scalars. In these models, the magnetic ordered phase only breaks the time reversal symmetry spontaneously. In a strict sense, if a field or operator $\mathcal{O}$ is invariant under the time reversal transformation, there is no interaction such as $B\mathcal{O}$ with respect to the external magnetic field $B$.

Note that a holographical antiferromagnetic phase is proposed by breaking a global SU(2) symmetry representing spin into a U(1) subgroup in a charged black hole background in Ref.~\cite{Iqbal:2010eh}. The symmetry breaking is triggered by condensation of a triplet scalar field charged under the SU(2) gauge field. Though SO(3) symmetry can be transformed into SU(2) symmetry and this model leads to the spatial rotational symmetry breaking spontaneously, the time reversal symmetry is not broken spontaneously in the magnetic ordered phase.  On the other hand, the attention in~\cite{Iqbal:2010eh} is focused on the symmetry breaking of paramagnetism/antiferromagnetism phase transition and the dispersion relation of the magnetic excitations. The structure of magnetic moment in material and the response to external magnetic field, which are the characteristic properties of antiferromagnetism, are not manifest in that model. So from the point of view of symmetry breaking and experimental phenomenological analysis,  Ref.~\cite{Iqbal:2010eh} seemingly does not give a complete model for paramagnetic/antiferromagentic phase transition.

Therefore a holographic model which is said to realize the phase transition of paramagnetism and antiferromagnetism should at least give three features: (1) The antiparallel magnetic moments structure as $T< T_N$. (2) The susceptibility behavior shown in figure~\ref{TM3}. (3) Breaking of time reversal symmetry, and of spatial rotational symmetry if spatial dimension is more than 2.  These are just what we want to realize in this paper. We will generalize the model in Ref.~\cite{Cai:2014oca} to describe the paramagnetism/antiferromagnetism phase transition in a dyonic black brane background by introducing two real antisymmetric tensor fields which couple to the background gauge field strength and interact with each other in the bulk. In the case without external magnetic field and in low temperatures, the magnetic moments of these two tensors condense spontaneously in the antiparallel manner with the same magnitude, which leads to an antiferromagnetic phase. Since these two magnetic moment both obtain minus signs under the  time reversal transformation, the time reversal symmetry is broken spontaneously. If the boundary spatial dimension is 3, magnetic moment is an axial vector and the spatial rotational symmetry is broken  spontaneously as well. In the case with weak external magnetic field, the magnetic susceptibility has a peak at the critical temperature and satisfies the Curie-Weiss law in the paramagnetic phase of antiferromagnetic materials. For strong external magnetic field, there is a critical magnetic field $B_c$. If $B>B_c$ the two magnetic moments in sublattices will have the same value paralleling to external magnetic field.

The paper is organized as follows. In section~\ref{model}, we will first introduce the model and derive its equations of motion (EoMs). Then we will compute the on shell Euclidean action of the model, which is equivalent to give the grand potential of dual boundary theory. The numerical calculations on the magnetic moment structure and the susceptibility in different phases will be given in section~\ref{numerc}. Because the calculations in the 3+1 bulk dimensional and 4+1 bulk dimensional cases are very similar, we will consider the 3+1 bulk dimensional case for simplification~\footnote{In AdS/CFT duality, 3+1 gravity is dual to a 2+1 boundary theory, which can be regarded as an effective description for some antiferromagnetic film in real materials.} and give some comments for the 4+1 dimensional case.  We will point out some differences and give some discussions in section~\ref{dim4} for the
4+1 dimensional case.  The summary and some discussions are included in section~\ref{summ}. A description in terms of mean field theory for the antiferromagnetism/paramagnetism phase transition is given in appendix~\ref{app1}.

\section{Holographic model}
\label{model}
\subsection{Model and EoMs}
Generalizing the model proposed in \cite{Cai:2014oca}, let us consider the following action:
\begin{equation}\label{action1}
S=\frac1{2\kappa^2}\int d^4x\sqrt{-g}[R+6/L^2-F^{\mu\nu}F_{\mu\nu}+\lambda^2(L_1+L_2+L_{12})],
\end{equation}
where
\begin{equation}\label{action2}
\begin{split}
&L_{(a)}=-\frac14\nabla^\mu M^{(a)\nu\tau}\nabla_\mu M^{(a)}_{\nu\tau}-\frac{1}4m^2 M^{(a)\mu\nu}M^{(a)}_{\mu\nu}\\ &-\frac12M^{(a)\mu\nu}F_{\mu\nu}-\frac {1}8 J V(M^{(a)}_{\mu\nu}),\\
&V(M^{(a)}_{\mu\nu})={M^{(a)\mu}}_{\nu}{M^{(a)\nu}}_{\tau}{M^{(a)\tau}}_{\sigma} {M^{(a)\sigma}}_{\mu},~a=1,2\\
&L_{12}=-\frac k2M^{(1)\mu\nu}M^{(2)}_{\mu\nu},
\end{split}
\end{equation}
$L$ is the radius of AdS space, $2\kappa^2=16\pi G$ with $G$  the Newton constant. In this model, $k$, $m^2$ and $J$ are all model parameters with $J<0$. $\lambda^2$ characterizes the back reaction of the two polarization fields $M^{(a)}_{\mu\nu}$ with $a=1,2$ to the background geometry, and $L_{12}$ describes
the interaction between two polarization fields.
When $k=0$, the two polarization fields decouple. In that case, the two polarization fields condense independently at the same critical temperature with the same magnitude.  However, when $k\neq0$, because of the interaction between them, resulting  magnetic moments are dependent. In some suitable range of the model parameters,  two condensed magnetic moments may appear in an antiparallel manner.

 Since two rank-two fields involve in this model, some remarks are in order. It is a well-known fact that it is quite difficult to write down a self-consistent high spin field theory in a flat spacetime (see \cite{Rahman:2013sta}, for example). Usually high spin field theory suffers from ghosts and causality violation. But such problems do not exist in this model and the ferromagnetic model in Ref. \cite{Cai:2014oca}. The reason is that our rank-two fields are two antisymmetric fields, which in fact are two massive 2-form fields rather than two spin-2 fields. The latter requires the field to be a symmetric and traceless one.  As a result, the spin degree in general dose not agree with the rank of the field and depends on the dynamic of the field~\cite{Duff:1999rk}. For example, the 2-form Kalb-Ramond field is spin 0 for massless case~\cite{Ogievetsky} and spin 1 for massive case~\cite{Kalb}. $p$-form field in string/M-theory is well defined, as the source of the $(p-1)$-brane. We can obtain a well defined theory for any form field localized in D-branes~\cite{Fu:2012sa}, which is entirely different from the case involving symmetric high rank tensor field. The spin of standard $p$-form field in type IIB supergravity theory can be found in the page 76 of Ref. \cite{Duff:1999rk}, where one can see that $p$-form field does not suffer from the troubles of high spin field. The mass of p-form field can be generated through Higgs mechanism, Stueckelberg mechanism, or topological mass generation method. As a result, the model here and the one proposed in Ref. \cite{Cai:2014oca} do not involve the higher spin fields and we do not worry the troubles such as ghost, causality violation and so on.

The equations of motion for polarization fields read
\begin{equation}\label{eomFM}
\begin{split}
&\nabla^2M^{(1)}_{\mu\nu}-m^2M^{(1)}_{\mu\nu}-kM^{(2)}_{\mu\nu}-J{M^{(1)}_\mu}^\delta {M^{(1)}_\delta}^\tau M^{(1)}_{\tau\nu}=F_{\mu\nu},\\
&\nabla^2M^{(2)}_{\mu\nu}-m^2M^{(2)}_{\mu\nu}-kM^{(1)}_{\mu\nu}-J{M^{(2)}_\mu}^\delta {M^{(2)}_\delta}^\tau M^{(2)}_{\tau\nu}=F_{\mu\nu}.
\end{split}
\end{equation}
In the probe limit, we can neglect the back reaction of the two polarization fields on the background geometry. The background we will consider is a dyonic Reissner-Nordstr\"om-AdS black brane solution of
the Einstein-Maxwell theory with a negative cosmological constant, and the metric reads~\cite{Cai:1996eg}
\begin{equation}\label{geom}
\begin{split}
  ds^2=r^2(-f(r)dt^2+dx^2+dy^2)+\frac{dr^2}{r^2f(r)},\\
   f(r)=1-\frac{1+\mu^2+B^2}{r^3}+\frac{\mu^2+B^2}{r^4}.
\end{split}
\end{equation}
Here both the black brane horizon and AdS radius have been set to be unity. The temperature of the black
brane is
\begin{equation}\label{Tem1}
T=\frac1{4\pi}(3-\mu^2-B^2).
\end{equation}
For the solution (\ref{geom}), the corresponding gauge potential is $ A_\mu=\mu(1-1/r)dt+Bx dy$. Here $\mu$
is the chemical potential and $B$ can be reviewed as the external magnetic field of dual boundary
field theory.

In the background (\ref{geom}), let us consider the dynamics of the two polarization fields. Following \cite{Cai:2014oca}, we take the ansatz for the polarization fields $M^{(a)}_{\mu\nu}$ as
\begin{equation}\label{Mcomp}
M^{(a)}_{\mu\nu}=-p^{(a)}(r)dt\wedge dr+\rho^{(a)}(r)dx\wedge dy,~~~a=1,2.
\end{equation}
For the sake of convenience for later discussions, we introduce two variables
\begin{equation}\label{ab1}
\alpha=\frac12(\rho^{(1)}+\rho^{(2)}),~~\beta=\frac12(\rho^{(1)}-\rho^{(2)}).
\end{equation}
Then we can use $\alpha $ and $\beta$ to describe the different magnetic phases. The paramagnetic, ferromagnetic, antiferromagnetic  and ferrimagnetic phases correspond to the cases of $\alpha=\beta=0$, $|\alpha|>|\beta|$, $|\alpha|=0,\beta\neq0$ and $0\neq|\alpha|<|\beta|$, respectively. In the end of this paper, we will discuss the ferrimagnetic phase briefly.

 In terms of variables~$\alpha$ and $\beta$, the equations of motion for $\rho^{(1)}$ and $\rho^{(2)}$ can be rewritten as
\begin{equation}\label{eqab}
\begin{split}
\alpha''+\frac{f'\alpha'}f+\frac{J\alpha^3}{4r^6f}+\left(\frac{3J\beta^2}{4r^6f}-\frac{2f'}{rf}-\frac4{r^2} -\frac{m^2+k}{r^2f}\right)\alpha&\\
=\frac{2B}{r^2f},&\\
\beta''+\frac{f'\beta'}f+\frac{J\beta^3}{4r^6f}+\left(\frac{3J\alpha^2}{4r^6f}-\frac{2f'}{rf}-\frac4{r^2} -\frac{m^2-k}{r^2f}\right)\beta&\\
=0.&
\end{split}
\end{equation}
As in~Ref.~\cite{Cai:2014oca}, our attention mainly focuses on the case near the critical temperature and we will ignore the dynamics of $p^{(1)}$ and $p^{(2)}$ since they are decoupled from equations of $\alpha$ and $\beta$. To solve the equations of motion (\ref{eqab}), we have to impose suitable boundary
conditions at the horizon and AdS boundary, respectively.
The two functions $\alpha$ and $\beta$ should be regular at the horizon, which implies
\begin{equation}\label{init1}
\begin{split}
&\alpha'=2\alpha-\frac{J\alpha(\alpha^2+3\beta^2)-4\alpha(m^2+k)-8B}{16\pi T},\\
&\beta'=2\beta-\frac{J\beta(\beta^2+3\alpha^2)-4\beta(m^2-k)}{16\pi T},
\end{split}
\end{equation}
at the horizon.
So once the initial values of $\alpha$ and $\beta$ are given at the horizon, one can integrate the equations~\eqref{eqab} to get out the solution. The asymptotic behavior of the solution should not change
the asymptotically AdS behavior of the background geometry. As a result, near the AdS boundary we can neglect the nonlinear terms in (\ref{eqab}) since the condensations for $\alpha$ and $\beta$ are small near the critical temperature, and obtain
\begin{equation}\label{eqab2}
\begin{split}
&\alpha''-\left(\frac4{r^2} +\frac{m^2+k}{r^2}\right)\alpha-\frac{2B}{r^2}=0,\\
&\beta''-\left(\frac4{r^2}+\frac{m^2-k}{r^2}\right)\beta=0.
\end{split}
\end{equation}
The behavior of the solutions of equations depends on the value of $m^2+k+4$. When $m^2+k+4=0$, the asymptotic solutions will have a logarithmic term, we will not consider this case in the present paper.  When $m^2+k+4\neq0$, we have the solution:
\begin{equation}\label{ab2}
\begin{split}
&\alpha=\alpha_+r^{(1+\delta_1)/2}+\alpha_-r^{(1-\delta_1)/2}-\frac{2B}{m^2+k+4},\\
&\beta=\beta_+r^{(1+\delta_2)/2}+\beta_-r^{(1-\delta_2)/2},\\
&\delta_1=\sqrt{17+4k+4m^2},~~\delta_2=\sqrt{17-4k+4m^2},
\end{split}
\end{equation}
where $\alpha_{\pm}$ and $\beta_{\pm}$ are all finite constants.
 To satisfy the Breitenlohner-Freedman (BF) bound in $AdS_4$ space, the two model parameters $m$ and $k$ have to obey: $4|k|<17+4m^2$.  On the other hand, as  discussed in Ref.~\cite{Cai:2014oca}, in order to make the probe limit and the on shell action well defined, we should take
\begin{equation}\label{shoot1}
\delta_1>1,~~ \delta_2>1, ~~\alpha_+=\beta_+=0.
\end{equation}
This can be understood in another way: when $B=0$, the constants $\alpha_+$ and $\beta_+$ should be viewed as the sources of the corresponding operators in the boundary field theory, according to the AdS/CFT duality. In order the symmetry to be broken spontaneously, one has to turn off the source terms.  In addition, in order to make the $\rho^{(1)}$ and $ \rho^{(2)}$ condense and to realize the antiferromagnetic phase in the case without external magnetic field when the temperature is low enough, the model parameters should violet the BF bound of $\beta$ on $AdS_2$, but not for $\alpha$,  which leads to
\begin{equation}\label{Ads2BF}
3+2(m^2+k)>0,~~3+2(m^2-k)<0.
\end{equation}
Considering \eqref{shoot1} and \eqref{Ads2BF} together, we see that $m^2$ and $k$ should satisfy
\begin{equation}\label{canshu}
\left|\frac32+m^2\right|<k<4+m^2.
\end{equation}

\subsection{On shell action and free energy}
\label{action}
To see the phase transition of the model, we should calculate the free energy of the dual field theory, which is given by the on shell Euclidean action of the gravity theory. Since we are working in the probe limit, we need only calculate the part of the two polarization fields in the dyonic black brane background.
 A direct computation gives:
\begin{equation}\label{onshellS}
\begin{split}
&S=S^{(1)}+S^{(2)},\\
&S^{(a)}=\lambda^2\int_{r\rightarrow\infty} d^3x\sqrt{-h}(-\frac{1}4n^\mu M^{(a)}_{\nu\tau}\nabla_{\mu}M^{(a)\nu\tau})\\
&+\frac{\lambda^2}4\int d^4x\sqrt{-g}[JV(M^{(a)}_{\mu\nu})-F^{\mu\nu}M^{(a)}_{\mu\nu}],
\end{split}
\end{equation}
where $h$ is the determinate of the induced metric at the AdS boundary and $n^{\mu}$ is the unit normal to the boundary.  With the asymptotic behavior of the solution at $r\rightarrow\infty$ and the restriction \eqref{canshu}, we find that the boundary term in~\eqref{onshellS} vanishes. As a result, the free energy density for the part of $\rho^{(a)}$ is
\begin{equation}\label{Grho}
\begin{split}
G_\rho&=\lambda^2\int_1^\infty dr\left[\frac{B\alpha}{r^2}+\frac{J(\rho^{(1)4}+\rho^{(2)4})}{4r^6}\right]\\
&=\lambda^2\left[\int_1^\infty dr\frac{J(\alpha^4+\beta^4+6\alpha^2\beta^2)}{2r^6}-BN\right].
\end{split}
\end{equation}
Here the total magnetic moment density is defined by
\begin{equation}\label{totN}
N=N_1+N_2=-\lambda^2\int_1^{\infty}\frac{\alpha}{r^2}dr,
\end{equation}
with
\begin{equation}\label{totN2}
N_a=-\frac{\lambda^2}2\int_1^{\infty}\frac{\rho^{(a)}}{r^2}dr, ~a=1,2.
\end{equation}
Here $N_1$ and $N_2$ can be interpreted as the magnetic moments of two sublattices in the boundary. We can see that once the condensation happens, the nontrivial solution is always more stable than the trivial solution without condensation, due to the fact that $J<0$ and $N>0$ as we will see shortly.

The order parameter for antiferromagnetic phase transition, i.e., the staggered magnetization, is defined as,
\begin{equation}
N^\dag=N_1-N_2=-\lambda^2\int_1^{\infty}\frac{\beta}{r^2}dr.
\end{equation}
Because the free energy~\eqref{Grho} needs to be invariant and magnetic field $B$ will be changed into $-B$ under the time reversal transformation, we can get the rules for magnetic moment and staggered magnetization under the  time reversal transformation, which is,
\begin{equation}\label{Ttrnas}
\rho^{(a)}\rightarrow-\rho^{(a)},~~~N^\dag\rightarrow-N^\dag.
\end{equation}
So if $N^\dag\neq0$, then the time reversal symmetry is broken spontaneously. One should note that if the boundary is 3+1 dimensional, magnetic components of polarization fields form two SO(3) axial vectors and staggered magnetization is also SO(3) axial vector. Then if $N^\dag\neq0$, the time reversal and spatial rotational symmetries are broken spontaneously and simultaneously. This case will be further discussed in section~\ref{dim4}.

\section{Phase transition and susceptibility}
\label{numerc}

Now let us solve the equations \eqref{eqab} under the conditions \eqref{shoot1} and~\eqref{canshu}. In what follows, we will take model parameters $m^2=-2,k=1$ as a typical example and work in grand canonical ensemble where the chemical potential is fixed. First, we will change the chemical potential~$\mu$ to get different temperature of the black brane for any given magnetic field~$B$. Second, we find out the suitable initial values of $\alpha$ and $\beta$ at the horizon to satisfy the boundary conditions~\eqref{shoot1} through the shooting method. Third, note that the system has the following scaling symmetry:
\begin{equation}\label{scaling1}
\begin{split}
&r\rightarrow ar,~~(t,x,y)\rightarrow a^{-1}(t,x,y),~~T\rightarrow aT,~\mu\rightarrow a\mu,\\
&\rho^{(a)}\rightarrow a^2\rho^{(a)},~B\rightarrow a^2B,~N\rightarrow aN,~\chi\rightarrow\chi/a.
\end{split}
\end{equation}
 Thus once obtain the solutions of \eqref{eqab}, we can use the scaling symmetry~\eqref{scaling1} to find
 the corresponding solutions with the same chemical potential.

\subsection{Critical temperature and phase transition}

\begin{figure}
\includegraphics[width=0.35\textwidth]{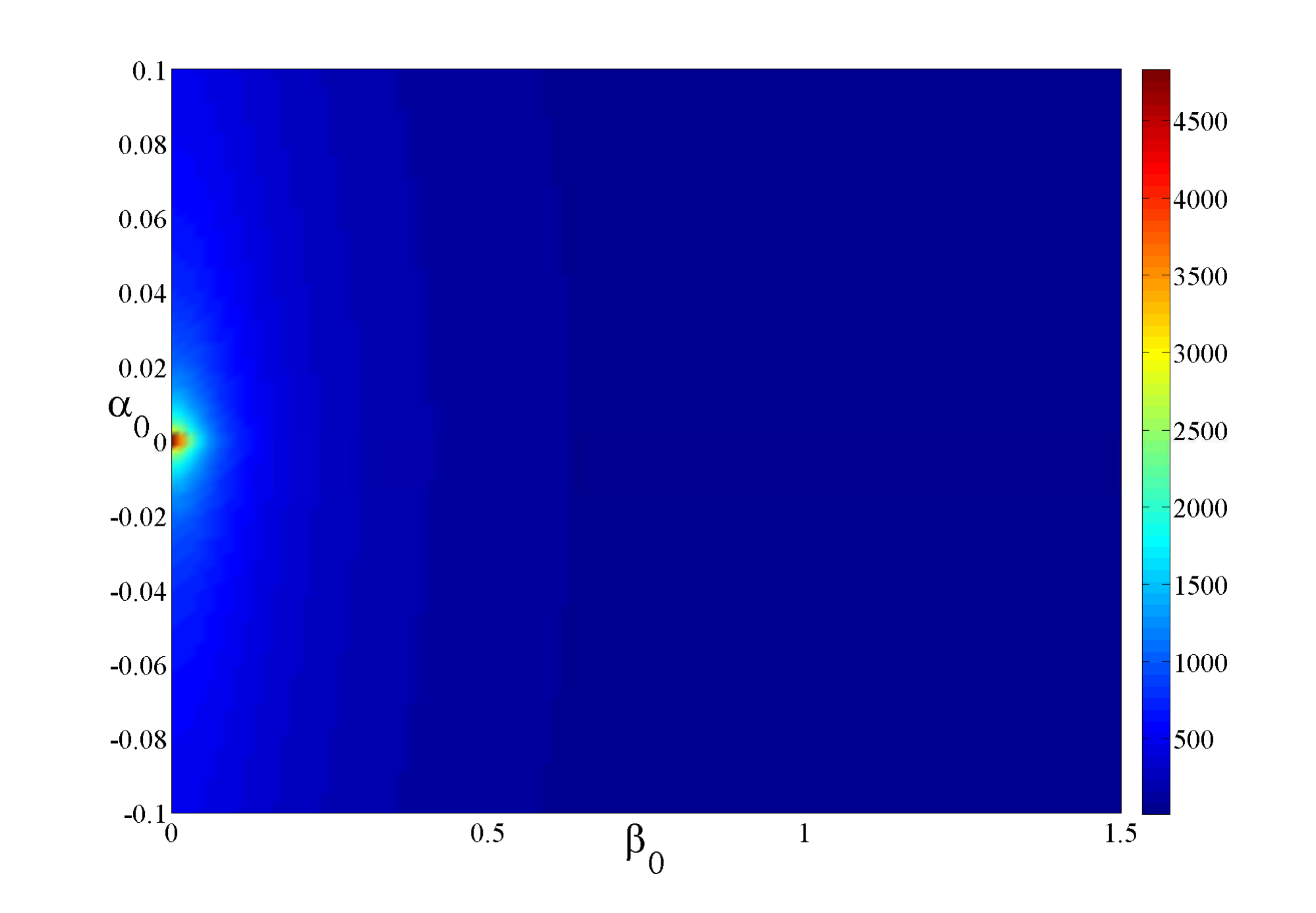}
\includegraphics[width=0.35\textwidth]{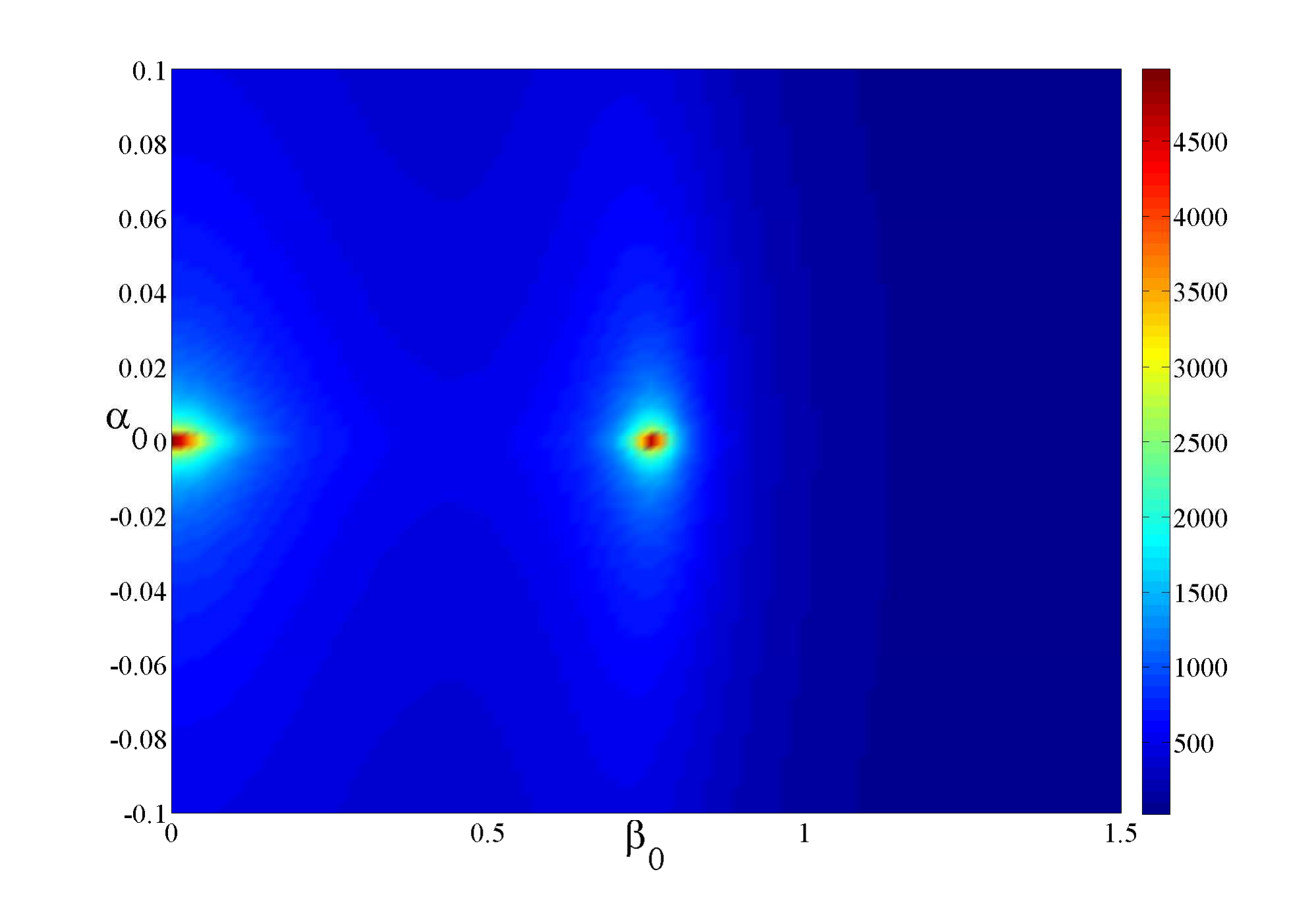}
\caption{The value of $(\alpha_+^2+\beta_+^2)^{-1}$ in the plane of $\alpha_0$-$\beta_0$ in different temperature. Top panel:~$T/\mu\simeq0.0111$. Bottom panel:~$T/\mu\simeq0.0083$.}
\label{Tgamma1}
\end{figure}

Let us first examine whether the polarization fields can condense in the antiferromagnetic manner when $B=0$. In order to do this, we will compute the values of $(\alpha_+^2+\beta_+^2)^{-1}$ for the different initial values $\alpha_0$ and $\beta_0$ at the horizon and plot them in the plane of $\alpha_0$-$\beta_0$. Each singularity in this plane corresponds to a solution of~\eqref{eqab} satisfying the condition~\eqref{shoot1}. Because of symmetry, we only need to check the region of $\beta_0>0$. The figure~\ref{Tgamma1} shows a typical example in the high and low temperature cases, respectively. In the high temperature case, we see that there is only one trivial solution locating at original point, which corresponds to a paramagnetic phase with $\alpha(r)=\beta(r)=0$. When the temperature is  low enough, we find there exist two nontrivial solutions which locate at~$\beta_0\simeq\pm0.754$ and $\alpha_0\simeq0$, respectively. They lead to the solution with~$\beta(r)\neq0,~\alpha(r)=0$, which corresponds to an antiferromagnetic phase.

 Thus we see that the model indeed can give rise to a paramagnetism/antiferromagnetism phase transition in the case without external magnetic field.  Since near the critical temperature, the value of $\beta(r)$ is very small and $\alpha(r)=0$, we can find the critical temperature by solving the linearized equation for $\beta$:
\begin{equation}\label{ct1}
\beta''+\frac{f'\beta'}f-\left(\frac{2f'}{rf}+\frac4{r^2} +\frac{m^2-k}{r^2f}\right)\beta=0.
\end{equation}
Solving equation~\eqref{ct1} with the condition of~$\beta_+=0$ by shooting method, we find the critical temperature is $T_N/\mu\simeq0.00925$ for the case with chosen model parameters.

Now we can conclude that the system is in the paramagnetic phase with magnetic disordered when the temperature is higher than $T_N$ and in the antiferromagnetic phase with two opposite magnetic moments when the temperature is lower than $T_N$. At $T_N$, there is a paramagnetism/antiferromagnetism phase transition. In addition, since the magnetic moments of the two polarization fields obtains expectation value, the time reversal symmetry is broken spontaneously. The numerical results shows that this phase transition is a second order one with $N^\dag= N_1-N_2\propto\sqrt{1-T/T_N}$, where $N_1$ and $N_2$ are the resulting magnetic moments of the two polarization field condensations, respectively. As discussed in the end of subsection~\ref{action}, this corresponds to a time reversal symmetry breaking spontaneously. In addition, let us stress here that if the boundary spatial dimension is three, the spatial rotational symmetry is also broken spontaneously, since the nonvanishing magnetic moment chooses a direction as special. These results are consistent with the mean field theory description of the paramagnetism/antiferromagnetism phase transition given in appendix~\ref{app1}.

\subsection{Static susceptibility in the paramagnetic and antiferromagnetic phases}

 Two opposite oriented magnetic moments are one of the characteristic properties of antiferromagnetic material. Another remarkable one is the behavior of susceptibility density of the material in the external magnetic field.  The static magnetic susceptibility density is defined by
%
\begin{equation}\label{chid}
\chi=\lim_{B\rightarrow0}\frac{\partial N}{\partial B}.
\end{equation}
When we turn on the external magnetic field $B$, the functions~$\alpha$ and~$\beta$ are both nonzero in any temperature. In order to compute the susceptibility density defined by~\eqref{chid}, we need to shoot for  the boundary conditions~\eqref{shoot1} with two parameters for equations~\eqref{eqab} under the given external magnetic field $B$. This make us have to find the zero points of a nonlinear function with two variables. However, since by definition, $\chi$ involves only the behavior of $B\rightarrow0$, the problem can be simplified in the following way.

In the case of $T>T_N$, taking the result~$\alpha=\beta=0$ when $B=0$ into account,  we expect that $\alpha(r)$ is in the same order as $B$, while $\beta$ is in the order as $B^2$. So in the case with weak magnetic field, we need not consider the equation for $\beta$ and can neglect the nonlinear terms in the equation of $\alpha$. In that case, we just need to solve the linear equation:
\begin{equation}\label{chi3}
\alpha''+\frac{f'\alpha'}f-\left(\frac{2f'}{rf}+\frac4{r^2} +\frac{m^2+k}{r^2f}\right)\alpha-\frac{2B}{r^2f}=0.
\end{equation}
Because $\alpha \propto B$, we can set $B=1$ simply. Solving equation~\eqref{chi3} with the boundary condition~$\alpha_+=0$, the susceptibility density then can be computed by
\begin{equation}\label{chi4}
\chi=N=-\lambda^2\int_1^{\infty}\frac{\alpha}{r^2}dr.
\end{equation}

 In the case of $T<T_N$, the system is in the antiferromagnetic phase. When $B=0$, the solution of~\eqref{eqab} has $\alpha=0,~\beta\neq0$. When turn on a small $B$, we  expect that $\alpha(r)$ is still in the same order as $B$, which leads to a term with the order of $B^2$ in the equation of $\beta$. However, since the calculation of the susceptibility density only involves the linear term with $B$, thus  equations~\eqref{eqab} can be approximated reasonably to
\begin{equation}\label{eqab3a}
\beta''+\frac{f'\beta'}f+\frac{J\beta^3}{4r^6f}-\left(\frac{2f'}{rf}-\frac4{r^2} +\frac{m^2-k}{r^2f}\right)\beta=0,
\end{equation}
and
\begin{equation}\label{eqab3b}
\alpha''+\frac{f'\alpha'}f+\left(\frac{3J\beta^2}{4r^6f}-\frac{2f'}{rf}-\frac4{r^2} -\frac{m^2+k}{r^2f}\right)\alpha=\frac{2B}{r^2f}.
\end{equation}
 In that case, the equation of $\beta$ is decoupled from $\alpha$. Thus we can solve $\beta$ firstly and then put it into equation~\eqref{eqab3b} and obtain $\alpha$ by solving (\ref{eqab3b}).  We can set~$B=1$
 in (\ref{eqab3b}) once again because $\alpha \propto B$. Then the susceptibility density can be computed also through \eqref{chi4}, once we have the solution $\alpha (r)$.

\begin{figure}[h!]
\includegraphics[width=0.35\textwidth]{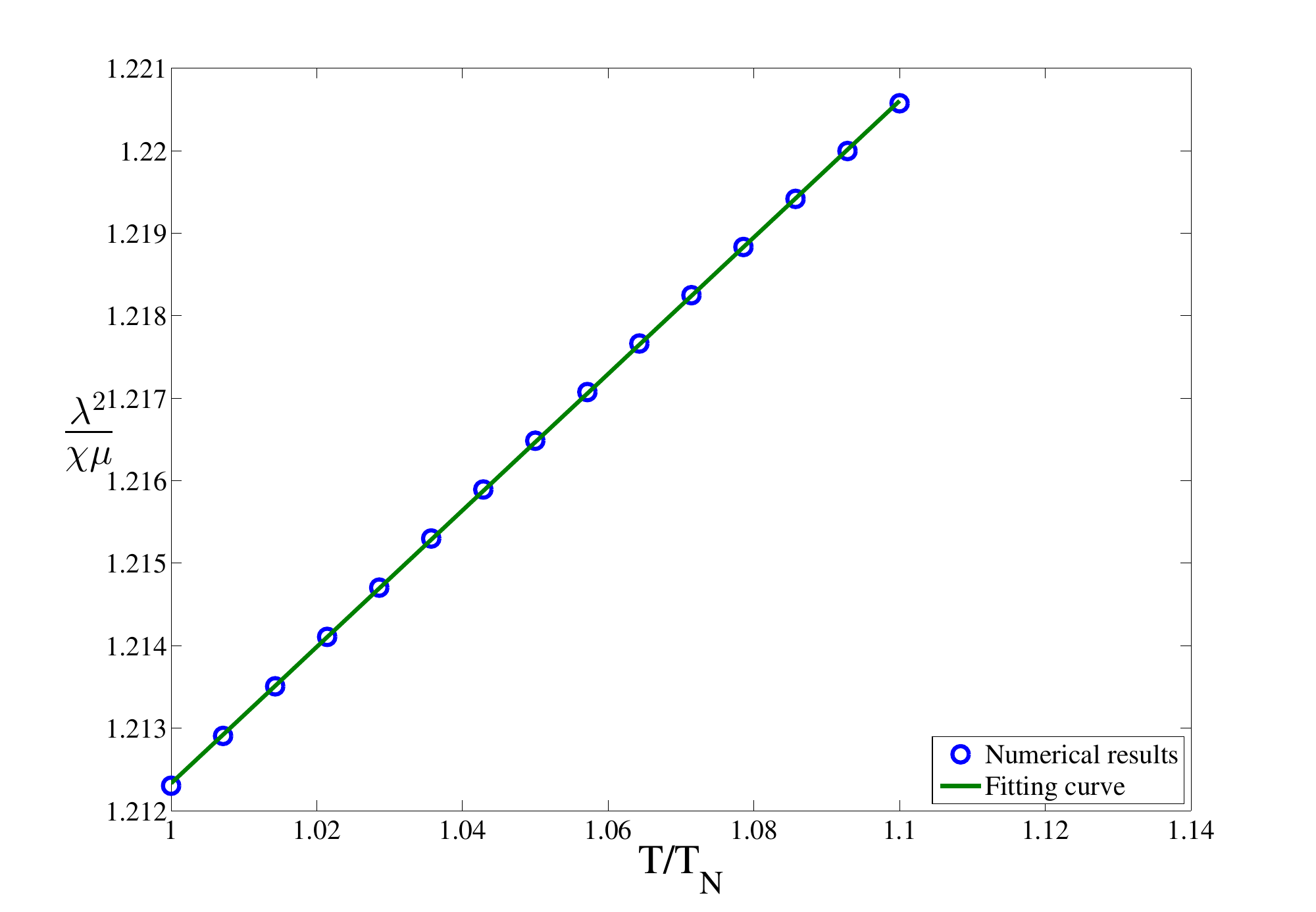}
\includegraphics[width=0.35\textwidth]{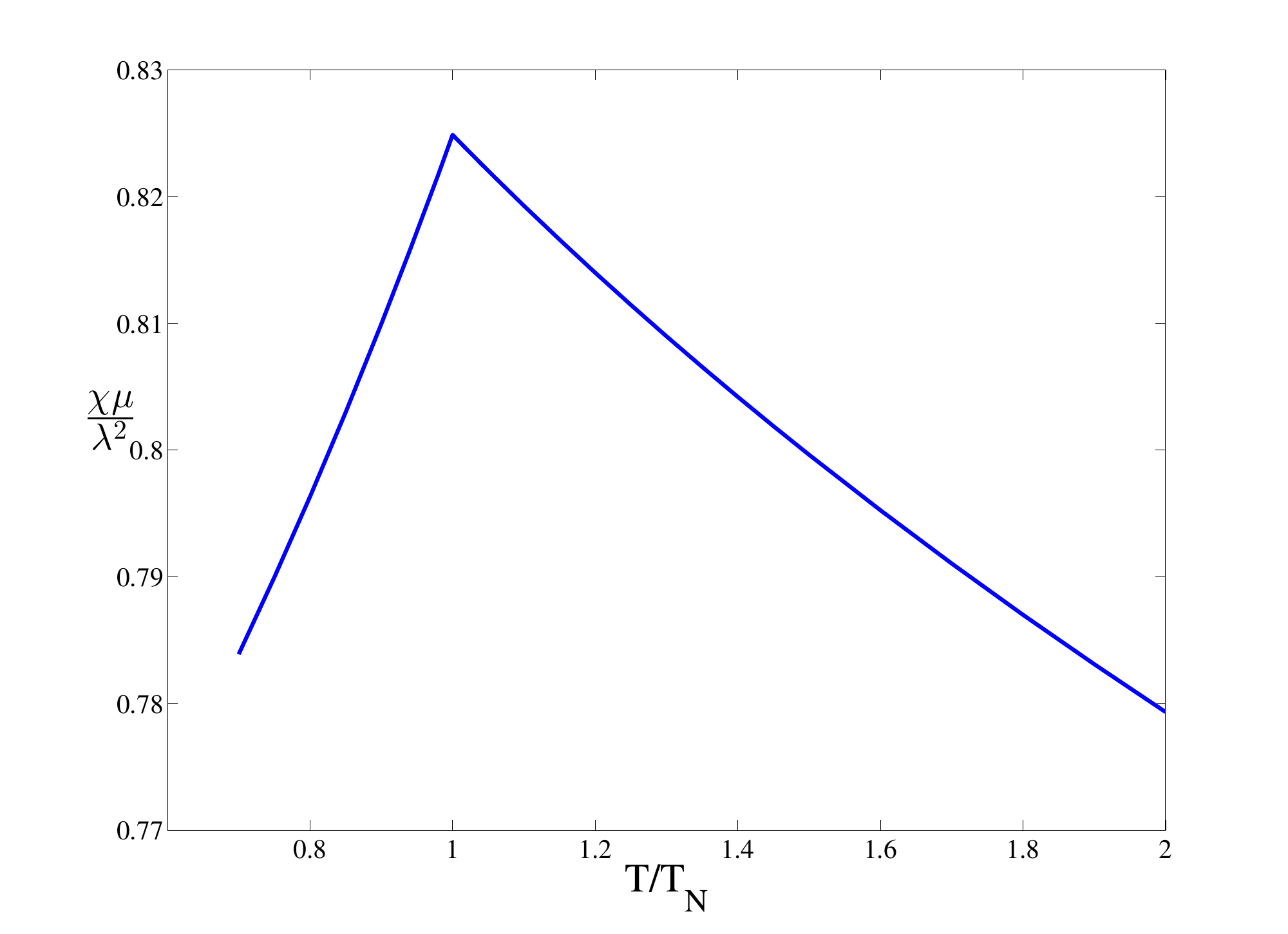}
\caption{The susceptibility density versus the temperature. (Top panel) The behavior of the inverse susceptibility density in the paramagnetic phase near the critical temperature. (Bottom panel) The susceptibility density in the antiferromagnetic phase and paramagnetic phase.}
\label{Tchi}
\end{figure}
Figure~\ref{Tchi} shows the behavior of the susceptibility density near the critical temperature $T_N$. In the paramagnetic phase, we can see that the susceptibility density increases when the temperature is lowered. Near the critical temperature, the susceptibility density satisfies the Curie-Weiss law of the antiferromagnetism~\eqref{chi1}. Concretely, for the chosen model parameters, we have
\begin{equation}
\lambda^2/\mu \chi \approx 0.0827(T/T_N+13.65),
\end{equation}
where $\theta/\mu\approx 13.65T_N/\mu =0.1263$. This is consistent with (\ref{chi1}).
When $T<T_N$, the susceptibility density increases when the temperature
increases. The susceptibility density has a clear peak at the critical temperature. These results are in agreement with the ones  shown in figure~\ref{TM3} and with the results from the mean field theory discussed in appendix ~\ref{app1} qualitatively.

\subsection{Influence of strong external magnetic field}

As mentioned in section~\ref{intro}, under the temperature below N\'{e}el temperature $T_N$, once the antiferromagnetic materials are put in an external magnetic field, the system will show a net magnetism because the two kinds of magnetic moments cannot offset each other. If the external magnetic field is so strong such as $B>B_c$, the magnetic moment structures will perform as paramagnetic materials where both two magnetic moments in two sublattices are the same and parallel to the direction of external magnetic field. According to the definition on $\alpha$ and $\beta$ in~\eqref{ab1}, this phenomenon corresponds to the case of $\alpha\neq0,~\beta\neq0$ with the nonzero magnetic field $B<B_c$ and $\alpha\neq0,~\beta=0$ if $B>B_c$ in our model.

In order to check these, let us see equation~\eqref{eqab} again. For a finite magnetic field $B$, we can see that the equation for $\alpha$ is an inhomogeneous equation. So for any temperature, there is always a nontrivial solution $\alpha(r)\neq0$ which satisfies the condition of~\eqref{shoot1}. From equation~\eqref{eqab}, we can read off the effective mass of $\alpha$ and $\beta$ near the horizon as
\begin{equation}\label{Meff1}
\begin{split}
&\widetilde{m}_{\alpha}^2={m^2+k}-\frac{3J\beta^2}{4r^4}+2rf',\\
&\widetilde{m}_{\beta}^2={m^2-k}-\frac{3J\alpha^2}{4r^4}+2rf'.
\end{split}
\end{equation}
From these relations, we can see that the increasing of $\alpha$ will increase the effective mass of $\beta$, which will suppress the condensate of $\beta$. If $B$ is not very large, then $\alpha(r)$ is also not very large so that $\beta$ can still condense. Then  we can get solution of~\eqref{eqab}
\begin{equation}\label{Bneq01}
\alpha\neq0,~\beta\neq0,~~~\text{when} ~T<T_N,~~\text{and} ~0<B<B_c.
\end{equation}
However, there is a critical magnetic field $B_c$, at which the effective mass of $\beta$ becomes large enough so that the condensation of $\beta$ disappears.

\begin{figure}[h!]
\includegraphics[width=0.35\textwidth]{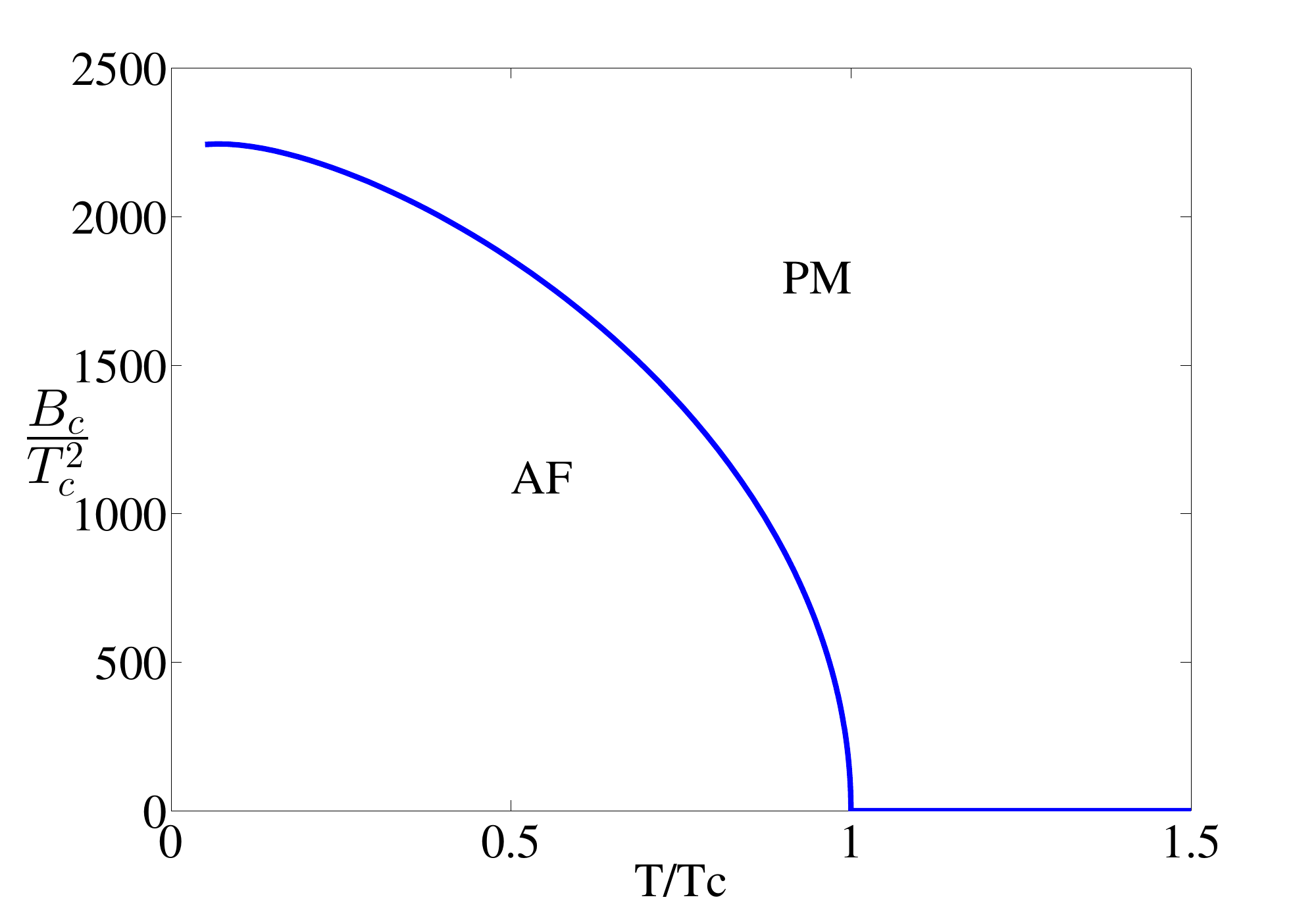}
\includegraphics[width=0.35\textwidth]{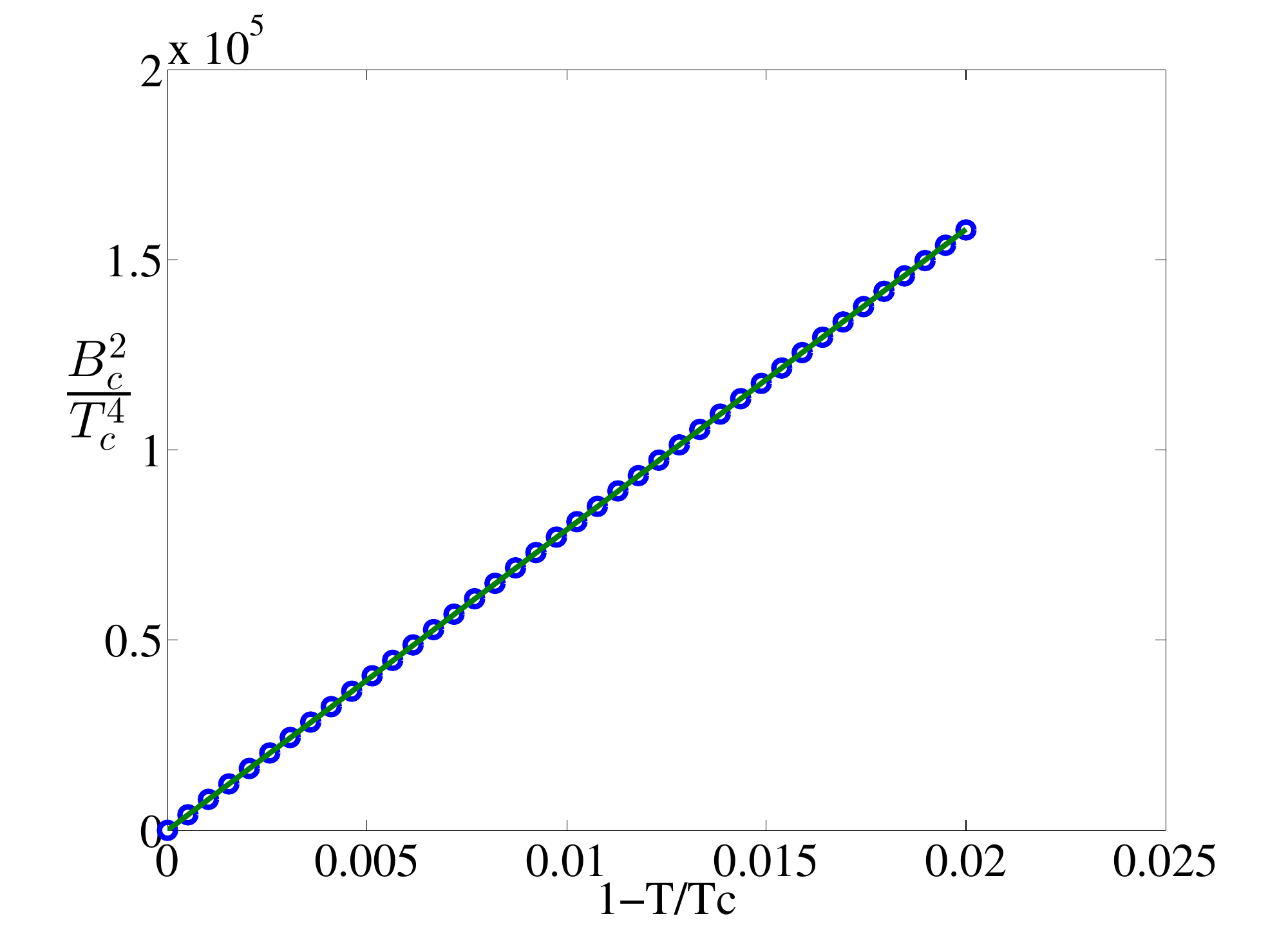}
\caption{The critical magnetic field versus the temperature.}
\label{TBc}
\end{figure}

Of course, the critical magnetic field $B_c$ depends on temperature $T$. In the zero temperature case, the $AdS_2$ BF-bound of $\beta$ is
\begin{equation}\label{BFbeta}
3+2(m^2-k-\frac34J\alpha_0^2)>0.
\end{equation}
Here $\alpha_0$ is the initial value of $\alpha$ at the horizon. when  the external magnetic field $B$ gets increased, then $\alpha_0$ will also increase. As the bound~\eqref{BFbeta} holds, the condensation of $\beta$ disappears. For the finite temperature case, the critical magnetic field can be obtained by solving the following equations
\begin{equation}\label{criticalB}
\begin{split}
&\alpha''+\frac{f'\alpha'}f+\frac{J\alpha^3}{4r^6f}-\left(\frac{2f'}{rf}+\frac4{r^2}+\frac{m^2+k}{r^2f}\right)\alpha+\frac{2B}{r^2f}=0,\\
&\beta''+\frac{f'\beta'}f+\left(\frac{3J\alpha^2}{4r^6f}-\frac{2f'}{rf}-\frac4{r^2} -\frac{m^2-k}{r^2f}\right)\beta=0.
\end{split}
\end{equation}
with the restrictions~\eqref{shoot1} and the initial condition $\beta|_{r_+=1}=1$. Figure~\ref{TBc} shows the behavior of critical magnetic field in different temperature. We can see that $B_c$ will increase when temperature decreases. This agrees with the physical expectation. When the temperature is lowered, the interaction between two magnetic moments is stronger and stronger. In order to destroy this interaction, the external magnetic field needs larger and larger.
Near the N\'{e}el temperature $T_N$, we can fit the behavior of $B_c^2$ with respect to $1-T/T_N$. The numerical results show a square root form
\begin{equation}\label{BcT}
B_c^2\simeq7.9\times10^6T_c^4(1-T/T_N).
\end{equation}
Because the staggered magnetization continuously decreases to zero when external field increases to $B_c$, there is a second order phase transition at $B=B_c$. One can confirm this second order phase transition by computing the free energy at the critical magnetic field and find that the free energy and its first order derivative are continuous at the critical point. The critical behavior near $B=B_c$ has been investigated recently in \cite{Cai:2015mja}.


\section{4+1 dimensional case}
\label{dim4}
In the previous section, our calculation is focused on the 3+1 bulk dimensional case, which corresponds to a 2+1 dimensional boundary theory.
In that case, only the nonvanishing magnetic moment comes from the component $M_{xy}$, such that the spatial rotational symmetry is in fact not broken even in the
condensed phase, although the time reversal symmetry is broken spontaneously.  Now we generalize the model into the 4+1 dimensional case in order to obtain a 3+1 dimensional boundary theory. In that case we can see clearly that the spatial rotational symmetry will be broken once the condensation happens.  For this, we only need change the geometry of~\eqref{geom} into the form in the 4+1 dimensional case. In the 4+1 dimensional case, the condition~\eqref{canshu} should be replaced by
\begin{equation}\label{canshu2}
\left|3+m^2\right|<k<6+m^2.
\end{equation}
In the 3+1 dimensional bulk case, the magnetic field is a pseudoscalar and the magnetic part in polarization field has only one component. The directions between these two components are parallel or antiparallel.  In the 4+1 dimensional bulk case, the components of magnetic parts of two polarization fields compose two SO(3) vectors, can this model also give two antiparallel magnetic moments rather than a non-collinear arrangement?  The answer is yes. A brief proof is given as follows.

In general, the ansatz for polarization fields can be written as
\begin{equation}\label{Mcomp2}
\begin{split}
M^{(a)}_{\mu\nu}=&-p^{(a)}(r)dt\wedge dr+\rho^{(a)}_z(r)dx\wedge dy+\rho^{(a)}_y(r)dz\wedge dx\\
&+\rho^{(a)}_x(r)dy\wedge dz,~~~a=1,2.
\end{split}
\end{equation}
When the external magnetic field is absent,  the equations of motion read
\begin{widetext}
\begin{equation}\label{eqM3d}
\rho''^{(a)}_{i}+\left(\frac{f'}f+\frac1r\right)\rho'^{(a)}_{i}+\frac{J\rho^{(a)2}}{r^6f}\rho^{(a)}_i-\left(\frac{m^2}{r^2f}+ \frac6{r^2}+\frac{2f'}{rf}\right)\rho^{(a)}_i-\frac{k\rho^{(b)}_i}{r^2f}=0,\\
\end{equation}
\end{widetext}
 where $(a,b)=(1,2)$ or (2,1) and $i=\{x,y,z\}$ and $\rho^{(a)2}=\rho^{(a)2}_z+\rho^{(a)2}_x+\rho^{(a)2}_y$. Because of the global SO(3) symmetry of the system, by adjusting the directions of axis \{x,y,z\}, we always can set following initial values at the horizon~\footnote{In fact, the global SO(3) symmetry permits us to set this condition at any point. But one cannot set this condition in an open interval or two different points in general.}
\begin{equation}\label{init2}
\rho^{(1)}_x(r_h)=\rho^{(1)}_y(r_h)=\rho^{(2)}_y(r_h)=0.
\end{equation}
By regular requirement at the horizon, the initial values of $\{\rho^{(a)}\}$ have following relations
\begin{equation}\label{init3}
\rho'^{(a)}_i=2\rho^{(a)}_i+\frac{\rho^{(b)}_i k+\rho^{(a)}m^2-J\rho^{(a)2}\rho^{(a)}_i}{4\pi T}.
\end{equation}
Combining the expressions~\eqref{init3} and~\eqref{init2}, we have $\rho^{(a)}_y(r_h)=\rho'^{(a)}_y(r_h)=0$ and only three free parameters at the horizon remain, which are,
\begin{equation}\label{free3}
\{\rho^{(1)}_z(r_h),~\rho^{(2)}_z(r_h),~\rho^{(2)}_x(r_h)\}.
\end{equation}
Once given the values of them, we can integrate the equations~\eqref{eqM3d} to the boundary to obtain the whole solution. But near the boundary, by the variable substitutions similar to~\eqref{ab1}, the linearized equations give following asymptotic solutions,
\begin{equation}\label{asy1}
\begin{split}
&\rho^{(1)}_i=\frac12(\alpha_i+\beta_i),~\rho^{(2)}_i=\frac12(\alpha_i-\beta_i),\\
&\alpha_i=\alpha_{i+}r^{\sqrt{6+m^2+k}}+\alpha_{i-}r^{\sqrt{6+m^2+k}},\\
&\beta_i=\beta_{i+}r^{\sqrt{6+m^2-k}}+\beta_{i-}r^{\sqrt{6+m^2-k}}.
\end{split}
\end{equation}
As mentioned in the previous section, we need to impose following boundary conditions,
\begin{equation}\label{condi}
\alpha_{i+}=\beta_{i+}=0,~~i=x,y,z,
\end{equation}
which give 6 boundary constraints. However, we  have only three free variables~\eqref{free3} at the horizon. As  a result, in general, there is no solution except for the case that $\rho^{(a)}_y(r)=\rho^{(a)}_x(r)=0$.  As to the remaining nontrivial components $\rho^{(a)}_z(r)$, as discussed in the previous section, under the restrictions~\eqref{canshu2}, they will condense with the same magnitude, but opposite signs.

Thus we conclude that our model can indeed realize the antiparallel magnetic moment structure in the magnetic ordered phase in the 4+1 dimensional bulk theory, which gives an expected  3+1 dimensional antiferromagnetic theory. Once the temperature is lower than a critical value, the system will enter into antiferromagnetic state with time reversal and spatial rotational symmetries broken spontaneously and simultaneously.

\section{Summary}
\label{summ}
In this paper, we have presented a holographic model to realize the paramagnetism~/antiferromagnetism phase transition in a dyonic black brane background by introducing two real antisymmetric polarization fields in the bulk. These two polarization fields couple to the background gauge field strength and interact with each other.  There exists a critical temperature $T_N$ in this model if the model parameters are in some suitable region.

In the case without the external magnetic field,  these two polarization fields are both zero and
the condensation does not happen in the high temperature region of $T>T_N$, which corresponds to a paramagnetic phase. However, when the temperature is lowered to the region of $T<T_N$,  the condensation
happens spontaneously and the resulting two magnetic moments appear in an antiparallel manner with the same magnitudes, which leads to an antiferromagnetic phase, where the time reversal and spatial rotational symmetries are broken spontaneously. As a result, this is a kind of paramagnetism/antiferromagnetism phase transition at the critical temperature $T=T_N$.

When a weak external magnetic field is turned on, the resulting magnetic moment of the system is always nonzero. We discussed its response to the external magnetic field by calculating the magnetic  susceptibility density. The results show that the magnetic susceptibility density has a peak at the critical temperature and satisfies the Curie-Weiss law in the paramagnetic phase, which agree with the properties of susceptibility density in a realistic antiferromagnetism.

Although most calculations are made in 3+1 bulk dimensions, our results can also be extended into the 4+1 bulk dimensional case in order to realize a 3+1 dimensional boundary theory. In details, in the 4+1 bulk dimensional case, we proved our model can realize the time reversal and spatial rotational symmetries broken spontaneously and simultaneously and the two magnetic moment vectors are indeed antiparallel.

This model can be generalized to describe the ferrimagnetism/paramagentim phase transition. Note that in the ferrimagnetic phase there exist two antiparallel magnetic moments, but with different magnitudes. This can be archived by making the two polarization fields have different masses, which lead to these two fields are not condensed at the same temperature,  or/and by considering different self-interacting constant $J$, which leads to the induced magnetic moments having different temperature-dependent behaviors. In such a model, we expect to have a complex spontaneous magnetization behavior and various susceptibility curves. We are going to study these issues in the future.

In appendix~\ref{app1}, we give a mean field theory description of the antiferromagnetic phase transition, which can be regarded as a comparison. One may wonder whether it is worth building a holographic model to describe such a phenomenon,  because it seemingly looks that what one can obtain from the holographic model can all be archived from a suitable mean field theory model. However, essentially these two
kinds of descriptions are different. First, the instability in the Landau-Ginzburg mean field theory must be put in by hand, while it arises
naturally in the holographic setup. Second, the Landau-Ginzburg model is only valid near the
transition point, whereas the gravitational description can characterize the whole dynamics.
For a given bulk action, scanning through values of model parameters corresponds to
scanning through many different dual field theories. In that sense, a simple holographic
model is of some universality, i.e., the results may be true for a large class of dual field
theories, and are quite insensitive to the details of their dynamics. Third, in a top-down holographic setup, in principle one can
know the details of dual field theory, while it is impossible in the mean field theory description, although our holographic model is
a bottom-up one.

 There are many critical phenomenons involving strong correlation in condensed matter physics, which are considered to be controlled by spontaneous magnetization or something relevant to it. The toy model presented in this paper can be regarded as a new starting point to investigate them from the holographic viewpoint.
 In particular, it gives us a wide possibility to investigate the coexistence, competition or interaction between magnetic orders and superconducting orders, by combing this model with the holographic superconductor model~\cite{Hartnoll:2008vx}. The highly possible and very important applications are to investigate the coexistence and competition in ferromagnetic superconductor or superconducting ferromagnetism~\cite{Karchev,MacHida,Nevidomskyy,Kordyuk} and the antiferromagnetic state in unconventional superconductor~\cite{Norman}.  All these works are expected to be realized in the future.

\section*{Acknowledgements}

This work was supported in part by the National Natural Science Foundation of China (No.10821504, No.11035008, No.11375247, and No.11435006), and in part by the Ministry of Science and Technology of China under Grant No.2010CB833004.

\appendix
\section{Mean field theory for antiferromagnetism}
\label{app1}
 In this appendix we give a brief review on the mean field theory description for antiferromagnetism. According to magnetic moment structure shown in figure~\ref{TM3}, one can introduce two magnetic moments $\overrightarrow{M}_1$ and $\overrightarrow{M}_2$. Near the critical temperature, the free energy can be assumed to be of the form
%
\begin{equation}\label{MN0}
\begin{split}
G=&G_0+\frac a2T(\overrightarrow{M}_1^2+\overrightarrow{M}_2^2)+kM_1M_2\cos\theta\\
&+\frac b4(\overrightarrow{M}_1^4+\overrightarrow{M}_2^4)-\overrightarrow{H}\cdot(\overrightarrow{M}_1+\overrightarrow{M}_2),
\end{split}
\end{equation}
%
where $a$, $k$ and $b$ are all positive constants, $\theta$ is the included angle between two magnetic moments. $\overrightarrow{H}$ is the external magnetic field, and $G_0$ stands for the irrelevant part. Comparing with the model of ferromagnetism/paramagnetism phase transition~\cite{R.M.W}, one can view this free energy as the one of two magnetic moments with interaction between them.  If  set $\overrightarrow{M}_1=\overrightarrow{M}_2$,  we then get a ferromagnetism/paramagnetism phase transition model with a negative critical temperature. Using the conditions of thermodynamic equilibrium, we have
\begin{equation}\label{MN1}
\left\{
\begin{split}
bM_1^3+aTM_1+kM_2\cos\theta-H\cos\phi_1&=0,\\
bM_2^3+aTM_2+kM_1\cos\theta-H\cos\phi_2&=0,\\
M_1M_2\sin\theta&=0.
\end{split}
\right.
\end{equation}
Here $\phi_1$ and $\phi_2$ are included angle of $\overrightarrow{H}$ and $\overrightarrow{M}_1$ and $\overrightarrow{M}_2$ respectively. At first, let us consider the case of $\overrightarrow{H}=0$. The solutions for the equations \eqref{MN1} are
\begin{equation}\label{MN2}
\left\{
\begin{split}
&\overrightarrow{M}_1=\overrightarrow{M}_2=0,~~~T>T_N=k/a,\\
&\overrightarrow{M}_1=-\overrightarrow{M}_2, ~~~|\overrightarrow{M}_1|=\sqrt{(1-T/T_N)k/b},~~T\leq T_N.
\end{split}
\right.
\end{equation}
Thus we can see that when the temperature crosses $T_N$ from the high temperature regime $T> T_N$, the system  becomes  magnetic ordered from  disordered with a pair of opposite orientated magnetic moments. As the magnetic moments condense in an opposite parallel manner, the material does not show any macroscopic magnetic moment.

\begin{figure}[h!]
\includegraphics[width=0.35\textwidth]{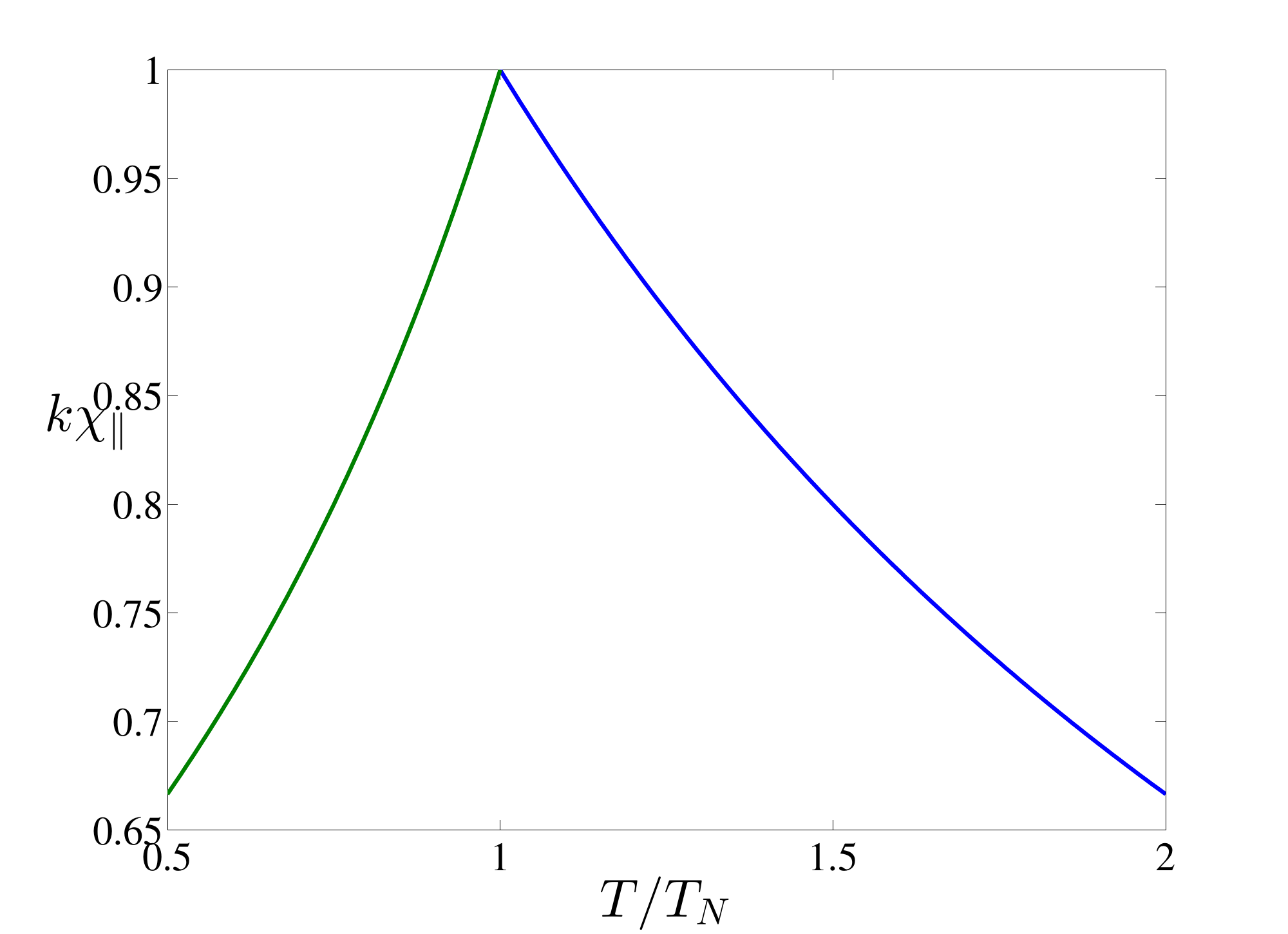}
\includegraphics[width=0.35\textwidth]{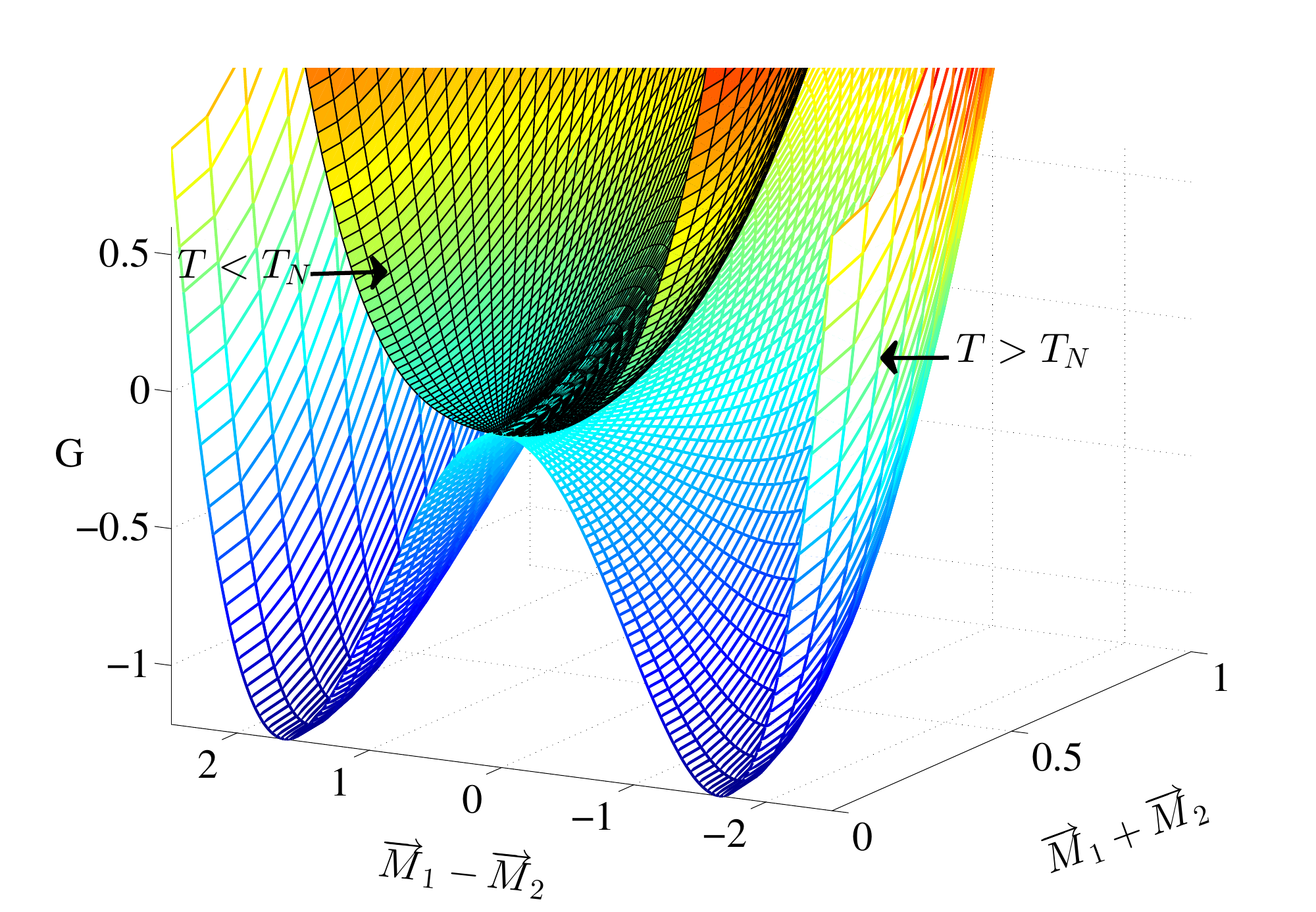}
\caption{(Top panel) The susceptibility versus temperature. We can see a peak at the critical temperature $T=T_N$.  (Bottom panel) The free energy of the model in the high temperature regime $T>T_N$ and low temperature regime $T<T_N$.}
\label{Tchi2}
\end{figure}

We  can compute the difference of free energy between the antiferromagnetic phase and paramagnetic phase when $T\leq T_N$. A direct calculation gives
\begin{equation}\label{free2}
\Delta G=-\frac{k^2(T/T_N-1)^2}{2b},~~~T\leq T_N.
\end{equation}
So when $T<T_N$, the antiferromagnetic phase is more stable than the paramagnetic phase.

When the external magnetic field appears, using equations~\eqref{MN1} and~solutions~\eqref{MN2}, we can obtain the magnetic susceptibility for the case with the two  magnetic moments being (anti-)parallel to the magnetic field as
\begin{equation}\label{chiT}
\chi_{\|}=\lim_{H\rightarrow0}(\frac{\partial M_1}{\partial H}+\frac{\partial M_2}{\partial H})=\left\{
\begin{split}
&\frac2{k(T/T_N+1)},~~~T>T_N,\\
&\frac1{k(2-T/T_N)},~~~T\leq T_N.
\end{split}
\right.
\end{equation}
In figure~\ref{Tchi2}, we plot the susceptibility~\eqref{chiT} and the free energy of the model in the case with $k=a=1,~b=1/3$ and~$G_0=0$. One can find that the susceptibility satisfies the Curie-Weiss law~\eqref{chi1} in high temperature regime $T>T_N$. At the critical temperature, the susceptibility has a peak. When $T<T_N$, the susceptibility decreases when the temperature lowers.  On the other hand, the free energy shows a  spontaneous breaking of time reversal and spatial rotational symmetries when $T<T_N$. As  a result the model of~\eqref{MN0} describes a second order phase transition and  the order parameter is the staggered magnetization $\overrightarrow{M}^\dag=\overrightarrow{M}_1-\overrightarrow{M}_2$.



\end{document}